\documentclass[11pt]{article}

\usepackage[colorlinks]{hyperref}
\hypersetup{
    colorlinks,	
    citecolor=blue,
}
\usepackage{color}

\begin{document}

\begin{titlepage}

\title{Regular Black Holes:\\
Towards a New Paradigm of Gravitational Collapse}

\author{Cosimo Bambi\footnote{Department of Physics, Fudan University, 200438 Shanghai, China; E-mail: bambi@fudan.edu.cn}  \hspace{0.05cm} (Editor)}

\maketitle

\vspace{0.5cm}

\begin{abstract}
Black holes are one of the most fascinating predictions of general relativity. They are the natural product of the complete gravitational collapse of matter and today we have a body of observational evidence supporting the existence of black holes in the Universe. However, general relativity predicts that at the center of black holes there are spacetime singularities, where predictability is lost and standard physics breaks down. It is widely believed that spacetime singularities are a symptom of the limitations of general relativity and must be solved within a theory of quantum gravity. Since we do not have yet any mature and reliable candidate for a quantum gravity theory, researchers have studied toy-models of singularity-free black holes and of singularity-free gravitational collapses in order to explore possible implications of the yet unknown theory of quantum gravity. This book reviews all main models of regular black holes and non-singular gravitational collapses proposed in the literature, and discuss the theoretical and observational implications of these scenarios.
\end{abstract}

\end{titlepage}

\section*{Preface}

General Relativity is one of the pillars of modern physics. While all the available observational data are in good agreement with the predictions of Einstein's gravity, there are a few unresolved theoretical issues that strongly point out the existence of new physics. One of the most outstanding and longstanding problems in General Relativity is the inevitability of spacetime singularities in physically relevant solutions of the Einstein Equations. At a spacetime singularity, predictability is lost and standard physics breaks down. It is widely believed that the problem of spacetime singularities can be solved within a theory of quantum gravity. However, despite the significant efforts and progress in the past decades, we do not have yet a robust theory of quantum gravity.

In the absence of a robust theory of quantum gravity, we can investigate qualitatively different scenarios that can solve the problem of spacetime singularities and can be obtained by violating specific conditions or assumptions valid in General Relativity. In other words, instead of following the more traditional top-down approach, in which we start from a well-formulated theory and we derive its predictions, we can try to follow a bottom-up strategy, in which we consider plausible predictions for a certain phenomenon and we try to derive the required key-ingredients of the fundamental theory.

Within this spirit, in the past years there has been an increasing interest in the study of regular black holes and singularity-free gravitational collapse models. Actually, the first attempts in this direction are not new, but initially they did not attract much interest from the community, and only in the past 10~years this line of research has grown significantly. The construction of the first regular black hole was presented by James Bardeen in 1968 at the {\it 5th International Conference on Gravitation and the Theory of Relativity} in Tbilisi and the first quantum gravity-inspired model of singularity-free gravitational collapse was discussed by Valeri Frolov and Grigorii Vilkovisky in 1979 at the {\it 2nd Marcel Grossmann Meeting} in Trieste. In 1998, Eloy Ayon-Beato and Alberto Garcia showed that it is possible to obtain regular black hole solutions in General Relativity in the presence of a nonlinear electrodynamic field. It was only around 2012-2013 that regular black holes started becoming a hot topic, as we can clearly see from the rise of the number of citations per year of the first works as well as the rise of the number of new publications on the topic.

The idea to publish a book on regular black holes and singularity-free gravitational collapse models is motivated, in part, by the fact this line of research has significantly grown in the past 10~years and there are a number of quite interesting recent results, and, at the same time, by the absence of a review covering all main models proposed so far. I am very grateful to all contributed authors of this book, who accepted enthusiastically to be involved in this project and wrote high quality contributions. I hope that this book can help young researchers to enter this exiting line of research and can contribute to progress further in our understanding of black holes and gravitational collapse models.

\vspace{\baselineskip}
\begin{flushright}\noindent
Shanghai, January 2023 \hfill {\it Cosimo Bambi}\\
\end{flushright}

\newpage

\section*{Contents}

\vspace{0.5cm}

\parbox[t]{14cm}{
Chapter~1

Regular Rotating Black Holes and Solitons with the de Sitter/Phantom Interiors

Irina Dymnikova

} 

\vspace{0.5cm}

  \noindent\parbox[t]{19cm}{
Chapter~2

Regular Black Holes Sourced by Nonlinear Electrodynamics

Kirill A. Bronnikov

e-Print: \href{https://doi.org/10.48550/arXiv.2211.00743}{arXiv:2211.00743}  } 

\vspace{0.5cm}

  \noindent\parbox[t]{19cm}{
Chapter~3

How Strings Can Explain Regular Black Holes

Piero Nicolini

e-Print: \href{https://doi.org/10.48550/arXiv.2306.01480}{arXiv:2306.01480}  } 

\vspace{0.5cm}

  \noindent\parbox[t]{19cm}{
Chapter~4

Regular Black Holes from Higher-Derivative Effective Delta Sources

Breno L. Giacchini, Tibério de Paula Netto

e-Print: \href{https://doi.org/10.48550/arXiv.2307.12357}{arXiv:2307.12357}  } 

\vspace{0.5cm}

  \noindent\parbox[t]{19cm}{
Chapter~5

Black Holes in Asymptotically Safe Gravity and Beyond

Astrid Eichhorn, Aaron Held

e-Print: \href{https://doi.org/10.48550/arXiv.2212.09495}{arXiv:2212.09495}  } 

\vspace{0.5cm}

  \noindent\parbox[t]{19cm}{
Chapter~6

Regular Black Holes in Palatini Gravity

Gonzalo J. Olmo, Diego Rubiera-Garcia

e-Print: \href{https://doi.org/10.48550/arXiv.2209.05061}{arXiv:2209.05061}  } 

\vspace{0.5cm}

  \noindent\parbox[t]{19cm}{
Chapter~7

Regular Black Holes from Loop Quantum Gravity

Abhay Ashtekar, Javier Olmedo, Parampreet Singh

e-Print: \href{https://doi.org/10.48550/arXiv.2301.01309}{arXiv:2301.01309}  } 

\vspace{0.5cm}

  \noindent\parbox[t]{19cm}{
Chapter~8

Gravitational Vacuum Condensate Stars

Emil Mottola

e-Print: \href{https://doi.org/10.48550/arXiv.2302.09690}{arXiv:2302.09690}  } 

\vspace{0.5cm}

 \noindent\parbox[t]{19cm}{
Chapter~9

Singularity-Free Gravitational Collapse: From Regular Black Holes to Horizonless Objects

Raúl Carballo-Rubio, Francesco Di Filippo, Stefano Liberati, Matt Visser

e-Print: \href{https://doi.org/10.48550/arXiv.2302.00028}{arXiv:2302.00028}  } 

\vspace{0.5cm}

  \noindent\parbox[t]{19cm}{
Chapter~10

Stability Properties of Regular Black Holes

Alfio Bonanno, Frank Saueressig

e-Print: \href{https://doi.org/10.48550/arXiv.2211.09192}{arXiv:2211.09192}  } 

\vspace{0.5cm}

  \noindent\parbox[t]{19cm}{
Chapter~11

Regular Rotating Black Holes

Ramón Torres

e-Print: \href{https://doi.org/10.48550/arXiv.2208.12713}{arXiv:2208.12713}  } 

\vspace{0.5cm}

  \noindent\parbox[t]{19cm}{
Chapter~12

Semi-classical Dust Collapse and Regular Black Holes

Daniele Malafarina

e-Print: \href{https://doi.org/10.48550/arXiv.2209.11406}{arXiv:2209.11406}  } 

\vspace{0.5cm}

  \noindent\parbox[t]{19cm}{
Chapter~13

Gravitational Collapse with Torsion and Universe in a Black Hole

Nikodem Popławski

e-Print: \href{https://doi.org/10.48550/arXiv.2307.12190}{arXiv:2307.12190}  }

\end{document}